\documentclass[preprint,showpacs,showkeys] {revtex4}
\usepackage{amssymb}
\usepackage{amsmath}
\usepackage{graphicx}

\begin{document}

\title{Polydispersity Effects in the Dynamics and Stability of
Bubbling Flows}

\author{E. Salinas-Rodr\'{\i}guez}
\altaffiliation{Fellow of SNI, Mexico.}
\email{sabe@xanum.uam.mx}
\affiliation{Departamento I. P. H., Universidad
Aut\'{o}noma Metropolitana,
Iztapalapa. Apdo.\ Postal 55--534, 09340 M\'{e}xico, D. F., M\'{e}xico\\
Instituto de F\'{\i}sica. UNAM.
Apdo.\ Postal 20--364, 01000 M\'{e}xico, D. F., M\'{e}xico}

\author{R. F. Rodr\'{\i}guez}
\altaffiliation[Also at ]{FENOMEC. Fellow of SNI, Mexico.}
\affiliation{Instituto de F\'{\i}sica. Universidad Nacional
Aut\'{o}noma de M\'{e}xico.\\
Apdo. Postal 20-364, 01000 M\'{e}xico, D. F., M\'{e}xico.}

\author{J. M. Zamora}
\altaffiliation{Fellow of SNI, Mexico.}
\author{A. Soria}
\altaffiliation{Fellow of SNI, Mexico.}

\affiliation{Departamento I. P. H., Universidad
Aut\'{o}noma Metropolitana-Iztapalapa.
Apdo.\ Postal 55--534, 09340 M\'{e}xico, D. F., M\'{e}xico\\
Instituto de F\'{\i}sica. UNAM.
Apdo.\ Postal 20--364, 01000 M\'{e}xico, D. F., M\'{e}xico}

\date{}

\begin{abstract}
The occurrence of swarms of small bubbles in a variety of industrial systems
enhances their performance. However, the effects that size polydispersity
may produce on the stability of kinematic waves, the gain factor, mean
bubble velocity, kinematic and dynamic wave velocities is, to our knowledge,
not yet well established. We found that size polydispersity enhances the
stability of a bubble column by a factor of about $23\%$ as a function of
frequency and for a particular type of bubble column. In this way our model\
predicts effects that might be verified experimentally but this, however,
remain to be assessed. Our results reinforce the point of view advocated in
this work in the sense that a description of a bubble column based on the
concept of randomness of a bubble cloud and average properties of the fluid
motion, may be a useful approach that has not been exploited in engineering
systems.
\end{abstract}



\maketitle

\section{Introduction}

The theoretical description of multiphase flows is
essentially based on analyzing the reponse of a cloud of dispersed particles
of different size ranges in a fluid. These particles constitute dynamic
phases and hence a multiphase flow. A widely used multiphase system is a
bubble column which is a reactor where a discontinuous gas phase in the form
of bubbles, moves relative to a continuous phase. Bubble columns have a wide
range of applications in chemical industries, biotechnology or in nuclear
reactors \cite{sha}, \cite{lapin}, \cite{yutani1}, \cite{yutani2}, \cite%
{nucreac}. The transient behavior is important at the start-up of these
systems and its analysis is essential in order to characterize the dynamic
performance of the columns. Among the phenomena that occur in these systems
void wave propagation mechanisms are of great importance since many
transient and steady states are controlled by the propagation of these waves
and, in this sense among others, the dynamic characterization of multiphase
flows is essential for the prevention of instabilities.

The (in)stability of bubbly flows which are characterized by almost
uniformly sized bubbles, is usually described in terms of the propagation
properties of void fraction and pressure disturbances caused by natural or
imposed fluctuations of the rate of air supply \cite{buscar1}, \cite{buscar2}%
, \cite{lahey}. Bubble size, rise velocity, size distribution and liquid and
bubble velocity profile have a direct bearing on the performance of bubble
columns. However, most of the time the dispersion devices deliver a dispersed
phase with a given size distribution. The importance of the size distributions
is only scarcely evaluated, most of the time by direct empirical trials and its
influence on the global behavior has still to be studied. Actually, to our
knowledge and from the theoretical point of view, it has not been yet well
established whether the stability of the motion of a swarm of bubbles is different
for monodispersed or polydispersed bubble flows. The main objective of this work
is to investigate the effects that size polydispersity might produce on the
stability of a bubble column. We shall introduce the effect of polidispersity
through the drag force in the hydrodynamic equations, using a method based on
statistical concepts and on a point-force approximation \cite{tam}. As we shall
see below, the corrections on the drag force factor, $C_{D}^{P}(a)$, due to
polydispersity depend only on the first three moments of a given particle size
distribution and they also have an effect on several properties of kinematic waves.
In particular, we found that size polydispersity enhances the stability of void
waves by a factor which varies between $4.5-23\%$ as a function of frequency and for a
particular type of bubble column. In this way our model\ predicts effects
that might be verified experimentally but this, however, remain to be
assessed.

To this end the paper is organized as follows. In Section 2 we briefly
review a hydrodynamical model for bubbly fluids introduced by Biesheuvel and
Gorissen \cite{byg}. Next, in Section 3 we consider a dispersion of
spherical air bubbles of different radi in water and we calculate the effect
of size polidispersity on the gain factor, mean bubble velocity, kinematic
wave velocities as a function of void fraction, for different wave
frequencies.

\section{Equations of motion of a bubble dispersion}

In this section we summarize the main ideas and steps behind the
hydrodynamical model for bubbly fluids introduced in Ref. \cite{byg}. The
equations of motion for a swarm of bubbles in a bubble column have been
derived in the literature by using standard methods of kinetic theory to
average over an ensamble or realizations of the flow \cite{hirschfelder}, 
\cite{ryg}. In Ref. \cite{byg} a dispersion of equally sized air bubbles in
a water column where the bubbles are small enough to remain spherical
through the whole system, is considered. They assumed that the air can be
taken as an incompressible fluid where no mass transfer is allowed between
the bubbles and the water, which is assumed to be an incompressible
Newtonian liquid. The conservation equation for the mean number density of
the gas bubbles, $n$, and the conservation equation for the mean bubble
momentum, $\rho _{G}\mathbf{v}$ (Kelvin impulse), were obtained for this
system \cite{lighthill}, 
\begin{equation}
\frac{\partial n}{\partial t}+\mathbf{\nabla }_{\mathbf{x}}\cdot \left( n%
\mathbf{v}\right) =0  \label{1}
\end{equation}
\begin{eqnarray}
&&\frac{\partial }{\partial t}\left[ n\left( \frac{4}{3}\pi a^{3}\rho _{G}%
\mathbf{v+I}_{L}\right) \right] +\mathbf{\nabla }_{\mathbf{x}}\cdot \left[
n\left( \frac{4}{3}\pi a^{3}\rho _{G}\mathbf{v+I}_{L}\right) \right] 
\nonumber \\
&&-\mathbf{\nabla }_{\mathbf{x}}\cdot \left( \mathbb{T}_{G}+\mathbb{T}%
_{L}\right)  \nonumber \\
&=&n\mathbf{F}_{D}+n\frac{4}{3}\pi a^{3}(\rho _{L}-\rho _{G})\mathbf{g.}
\label{2}
\end{eqnarray}%
$\mathbf{I}_{L}$ is the fluid impulse, $\mathbb{T}_{L}\left( \mathbf{x}%
,t\right) $ and \ $\mathbb{T}_{G}\left( \mathbf{x},t\right) $ are the fluid
stresses; $\mathbf{F}_{D}$\ is the drag force exerted by the fluid on the
bubble and $\mathbf{g}$ stands for the gravity field; $\rho _{L},$ $\rho
_{G} $ denote, respectively, the mass densities of water and air. $\mu _{L}$
stands for the liquid's viscosity. In order to describe the flow parameters
of the bubble swarm, Eqs. (\ref{1}) and (\ref{2}) should be expressed in
terms of the volume fraction of bubbles (or void fraction) $\varepsilon $
and their velocity field $\mathbf{v}$. Following Ref. \cite{byg} we assume
that the uniform flow of bubbles is along the axial direction of the column
with a mean axial rise velocity $v_{0}(\varepsilon )$, Therefore, $%
\varepsilon (z,t)\equiv \frac{4}{3}\pi a^{3}n(z,t)$.

The effect of hydrodynamic interactions between the bubbles on the
mean frictional force may be represented by introducing a function $%
f_{0}(\varepsilon )$ into $v_{0}(\varepsilon )$ in the form $%
v_{0}(\varepsilon )=f_{0}^{-1}(\varepsilon )v_{\infty }$, \cite{byg}. The
magnitude of the terminal velocity, $v_{\infty }$, of a single bubble of
radius $a$ in a stagnant liquid is given by \cite{batchelor1} $v_{\infty
}\equiv C_{D}^{-1}(\rho _{L}-\rho _{G})g$, where $C_{D}\equiv $ $9\mu
_{L}/a^{2}$ is the drag force factor and experiments suggest that \cite%
{wallis}, $f_{0}(\varepsilon )=\left( 1-\varepsilon \right) ^{-2}$. \bigskip
 The mean fluid impulse is modelled by
\begin{equation}
nI_{L}=n\left( \frac{2}{3}\pi a^{3}\rho _{L}\right) m_{0}(\varepsilon
)v_{0}(\varepsilon ),  \label{7}
\end{equation}%
where $m_{0}(\varepsilon )$ takes into account the effect of the
hydrodynamic interactions. According to Ref. \cite{bys} an expression for $%
m_{0}(\varepsilon )$ that renders reliable results up to large values of $%
\varepsilon $ is $m_{0}(\varepsilon )=(1+2\varepsilon )/(1-\varepsilon )$.

Since in a nonuniform bubbly flow the stress $\mathbb{T}=\mathbb{T}_{G}+$ $%
\mathbb{T}_{L}$ play the role of an effective pressure, they also assume
that the \textit{kinetic}\ contribution, $p_{e}(\varepsilon )$, is
proportional to the effective density of the bubbles, $\varepsilon ^{-1}\rho
_{ef}(\epsilon )\equiv \rho _{G}+\frac{1}{2}\rho _{L}m_{0}(\varepsilon )$,
and to the mean square of their velocity fluctuations $\overline{\Delta v^{2}%
\text{ }}\equiv H(\varepsilon )v_{0}^{2}(\varepsilon )=\frac{\varepsilon }{%
\varepsilon _{cp}}\left( 1-\frac{\varepsilon }{\varepsilon _{cp}}\right)
v_{0}^{2}(\varepsilon )$, \cite{batchelor1}. Here $\varepsilon _{cp}$ stands
for the limit of closest packaging of a set of spheres and is close to the
value 0.62. Thus, $p_{e}(\varepsilon )=\rho _{ef}\overline{\Delta v^{2}\text{
}}$. Furthermore, if the non-uniformity is the main cause of an additional
transfer of bubble momentum and fluid impulse associated with stress,
Biesheuvel and Gorissen \cite{byg} postulate that such a contribution to the
stress should be given by the force $\mu _{e}(\varepsilon )\frac{\partial }{%
\partial z}v$. Therefore, taking into account both contributions to the
stress, $\mathbb{T=-}p_{e}(\varepsilon )+\mu _{e}(\varepsilon )\frac{%
\partial v}{\partial z}$, where $v$ is the one dimensional nonuniform flow
velocity and $\mu _{e}(\varepsilon )=a\rho _{ef}(\varepsilon
)v_{0}(\varepsilon )H^{1/2}(\varepsilon )$ is an effective viscosity.

On the other hand, the mean frictional force is enhanced by an effective
diffusive flux of bubbles due to their fluctuating motion. This effect is
similar to an steady drag force acting upon each one of the bubbles and
proportional to the mean number density gradient. Therefore, using (\ref{2})
this force is represented by $nF_{D}=C_{D}\varepsilon f_{0}(\varepsilon )[v+%
\frac{\mu _{e}(\varepsilon )}{\varepsilon }\frac{\partial \varepsilon }{%
\partial z}]$. Substitution of the above expressions into Eqs. (\ref{1}) and
(\ref{2}) leads to the following closed set of one-dimensional equations of
motion for the bubbly flow in a zero volume flux reference frame, 
\begin{equation}
\frac{\partial \varepsilon }{\partial t}+\frac{\partial }{\partial z}%
(\varepsilon v)=0,  \label{14}
\end{equation}
\begin{eqnarray}
&&\frac{\partial }{\partial t}\left[ \rho _{ef}(\varepsilon )v\right] +\frac{%
\partial }{\partial z}\left[ \rho _{ef}(\varepsilon )v^{2}\right] -\frac{%
\partial }{\partial z}\mathbb{T}  \nonumber \\
&=&-C_{D}\varepsilon f_{0}\left( v+\frac{\mu _{e}(\varepsilon )}{\varepsilon
\rho _{ef}}\frac{\partial \varepsilon }{\partial z}\right) -\varepsilon
\left( \rho _{G}-\rho _{L}\right) g.  \label{15}
\end{eqnarray}

These equations may be rewritten in a laboratory reference frame by
considering the mean axial velocity of the dispersion, $U$, defined by $%
U(t)\equiv \varepsilon U_{G}+\left( 1-\varepsilon \right) U_{L}$. Here $%
U_{G} $ and $U_{L}$ are the mean bubble and fluid axial velocity in the
laboratory reference frame. Note that due to the incompressibility of both,
liquid and gas, $U$ is only a function of time. Therefore $v\equiv U_{G}-U$
and a Galileo transformation of Eqs. (\ref{14}) and (\ref{15}) gives
\begin{equation}
\frac{\partial \varepsilon }{\partial t}+\frac{\partial }{\partial z}%
\varepsilon U_{G}=0,  \label{17}
\end{equation}
\begin{eqnarray}
&&\frac{\partial }{\partial t}\left[ \varepsilon \left( \rho _{G}U_{G}+\frac{%
1}{2}\rho _{L}m_{0}\left( U_{G}-U\right) \right) \right]  \nonumber \\
&&+\frac{\partial }{\partial z}\left[ \varepsilon \left( \rho _{G}U_{G}+%
\frac{1}{2}\rho _{L}m_{0}\left( U_{G}-U\right) \right) U_{G}\right] 
\nonumber \\
&&-\frac{\partial }{\partial z}\left( -p_{e}+\mu _{e}\frac{\partial U_{G}}{%
\partial z}\right) -\varepsilon \rho _{G}\frac{\partial U}{\partial t} 
\nonumber \\
&=&-C_{D}\varepsilon f_{0}\left( \left( U_{G}-U\right) +\frac{\mu
_{e}(\varepsilon )}{\varepsilon \rho _{ef}}\frac{\partial \varepsilon }{%
\partial z}\right) -\varepsilon \left( \rho _{G}-\rho _{L}\right) g,
\label{18}
\end{eqnarray}%
together with the incompressibility condition
\begin{equation}
\frac{\partial U}{\partial z}=0.  \label{19}
\end{equation}

Consider a quiscent equilibrium state of the dispersion described by $%
\varepsilon =\varepsilon _{0}$. The deviations from this state will be
denoted by $\delta \varepsilon (z,t)$ and $\delta v(z,t)$. Linearization of
Eqs. (\ref{17}) - (\ref{19}) around the reference state yields the
wave-hierarchy equation%
\begin{eqnarray}
&&\tau _{\varepsilon }\left[ \left( \frac{\partial }{\partial t}+c^{+}\frac{%
\partial }{\partial z}\right) \left( \frac{\partial }{\partial t}+c^{-}\frac{%
\partial }{\partial z}\right) \delta \varepsilon -\nu _{\varepsilon }\left( 
\frac{\partial }{\partial t}+U_{G_{0}}\frac{\partial }{\partial z}\right) 
\frac{\partial ^{2}\delta \varepsilon }{\partial z^{2}}\right]  \nonumber \\
&&=-\left[ \left( \frac{\partial }{\partial t}+c_{0}\frac{\partial }{%
\partial z}\right) \delta \varepsilon -\nu _{\varepsilon }\frac{\partial
^{2}\delta \varepsilon }{\partial z^{2}}\right]  \label{20}
\end{eqnarray}%
with lower and higher-order wave velocities given by $c_{0}\equiv
U_{G_{0}}+\varepsilon _{0}v_{0}^{\prime }$ and
\begin{equation}
c^{\pm }\equiv U_{G_{0}}-\frac{\frac{1}{4}\varepsilon _{0}\rho
_{L}v_{0}m_{0}^{\prime }}{\rho _{G}+\frac{1}{2}\rho _{L}m_{0}}\pm \left[
\left( \frac{\frac{1}{4}\varepsilon _{0}\rho _{L}v_{0}m_{0}^{\prime }}{\rho
_{G}+\frac{1}{2}\rho _{L}m_{0}}\right) ^{2}+\frac{p_{e}^{\prime }}{\rho _{G}+%
\frac{1}{2}\rho _{L}m_{0}}\right] ^{1/2}.  \label{22}
\end{equation}%
Here $\nu _{e}(\varepsilon )=v_{0}(\varepsilon )H^{1/2}(\varepsilon )$ and $%
\tau _{e}(\varepsilon )=(C_{D}f_{0})^{-1}\left[ \rho _{G}+\frac{1}{2}\rho
_{L}m_{0}(\varepsilon )\right] $. The primes ($\prime $) denote derivatives
with respect to $\varepsilon $ and evaluated at the unperturbed state $%
\varepsilon =\varepsilon _{0}$.

For relatively low radial frequencies the wave propagation is described by a
linearized Burgers/Korteweg-de Vries equation%
\begin{equation}
\left( \frac{\partial }{\partial t}+c_{0}\frac{\partial }{\partial z}\right)
\delta \varepsilon \approx \left[ \tau _{\varepsilon
}(c^{+}-c_{0})(c_{0}-c^{-})+\nu _{\varepsilon }\right] \frac{\partial
^{2}\delta \varepsilon }{\partial z^{2}}+\tau _{\varepsilon }\nu
_{\varepsilon }(U_{G_{0}}-c_{0})\frac{\partial ^{3}\delta \varepsilon }{%
\partial z^{3}},  \label{25}
\end{equation}%
with a solution $\varepsilon \propto \exp (\gamma z-i\omega t)$ where $%
\omega $ is the frequency of the void wave and
\begin{equation}
\gamma \approx \frac{i\omega }{c_{0}}\left[ 1-\frac{\nu _{\varepsilon }\tau
_{\varepsilon }\omega ^{2}\left( U_{G}-c_{0}\right) }{c_{0}^{3}}\right] -%
\frac{\nu _{\varepsilon }\omega ^{2}\left( U_{G}-c_{0}\right) }{c_{0}^{3}}%
\left[ \tau _{\varepsilon }(c^{+}-c_{0})(c_{0}-c^{-})+\delta _{\varepsilon }%
\right] .  \label{27}
\end{equation}%
In terms of these quantities the so called gain factor, $G_{f}\equiv \exp %
\left[ \text{Re}(\gamma )\omega ^{2}\Delta z\right] $, where $\text{Re}%
(\gamma )$ denotes the real part and $\Delta z$ the distance between two
impedance probes in the experiments to measure $G_{f}$ \cite{mercadier}.

\section{\protect Polydispersed dispersion}

 The method developed by Tam \cite{tam} uses the concept of
randomness of the bubble cloud and derives equations describing the average
properties of the fluid motion. These averages are taken over a statistical
ensemble of particle configurations. A slow viscous flow past a large
collection of spheres of a given size distribution is considered to derive a
particle drag formula free from empirical assumptions. The result
essentially replaces the disturbance produced by a sphere in low Reynolds
number flow, by that of a point force located at the centre of the sphere.
The correction drag force factor is given by 
\begin{equation}
C_{D}^{P}=\lambda C_{D}\equiv \left[ 1+\alpha \overline{a}+\frac{1}{3}%
(\alpha \overline{a})^{2}\right] C_{D},  \label{28}
\end{equation}%
where 
\begin{equation}
\alpha =\frac{6\pi M_{2}+\left[ \left( 6\pi M_{2}\right) ^{2}+12\pi
M_{1}(1-3c)\right] ^{1/2}}{(1-3c)}.  \label{29}
\end{equation}%
\ $M_{n}=\int n(a)a^{n}da$ are the moments of the size distribution 
$n(a)$ and $c\equiv \frac{4}{3}\pi M_{3}$.

Since the terminal velocity of a bubble
depends
on $C_{D}$, it is reasonable to assume that in the polydispersed case $%
v_{0}(\varepsilon )$ should be replaced by $%
v_{0}^{P}\equiv \lambda ^{-1}v_{0}$. Substitution of this asumption into
Eqs. (\ref{17}) - (\ref{19}), carrying out the linearization procedure
described in the last section and using the explicit expressions of $\beta
\equiv \left\{ U_{G_{0}},c_{0},c^{\pm },\tau _{\varepsilon }\right\} $, one
can show that these quantities scale as $\beta ^{p}\equiv \left\{ \lambda
^{-1}U_{G_{0}},\lambda ^{-1}c_{0},\lambda ^{-1}c^{\pm },\lambda ^{-1}\tau
_{\varepsilon }\right\} $. If these polydispersed quantities are substituted
into Eq. (\ref{27}), one obtains an expression for the polydispersed gain
factor $G_{f}^{p}\equiv \exp \left[ \text{Re}(\gamma ^{p})\omega ^{2}\Delta z%
\right] $.

\section{RESULTS}

To compare the monodisperesed and polydispersed results on the
gain factor, mean bubble velocity, kinematic wave velocities as a function
of void fraction, we used the the following parameter values for an
air-water bubble column, $\Delta z=20$ $cm$, $V_{T}=1280$ $cm^{3}$, $\rho
_{G}=1.2046$ $\times $ $10^{-3}$ $gr/cm^{3}$, $\rho _{L}=0.998$ $gr/cm^{3}$, 
$\mu _{L}=1.002$ $\times $ $10^{-2}poise$, $\varepsilon _{cp}=0.62$. In Fig.
1 we plot $G_{f}$ and $G_{f}^{p}$ vs. $\varepsilon $ for different
frequencies $\omega $ and for a log-normal distribution $n(a)$ with average
and dispersion $\overline{a}=0.04\ cm$, $\sigma =0.5$, respectively. 
\begin{figure}
\includegraphics{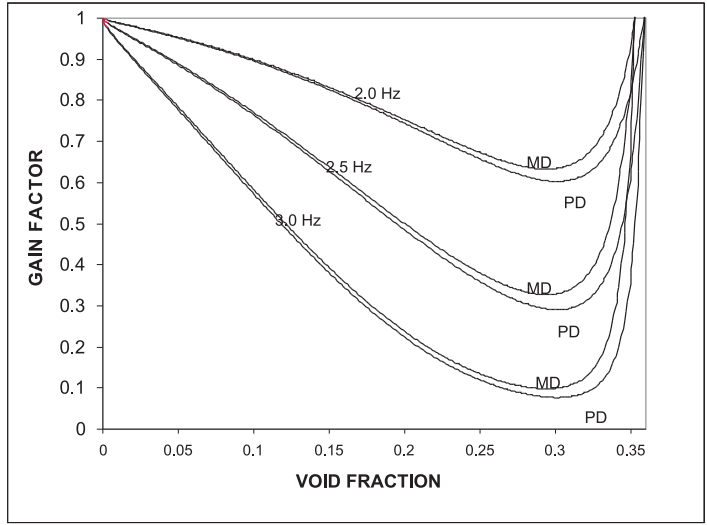}
\caption{Gain factors $G_{f}$, $G_{f}^{p}$ vs. $\varepsilon $
for waves with frequencies of $2$, $2.5$ and $3Hz$. The liquid is stagnant
and the parameter values are those given in section 4.}
\end{figure}

Note that for values $0.185\leqslant $ $%
\varepsilon \leqslant 0.301$ the attenuation rate drops significantly for
For instance,the per cent difference defined by $\Gamma
\equiv $ $\left\vert G_{f}-G_{f}^{p}\right\vert /G_{f} $. For the range
both cases. For instance, for a frequency of $2Hz$ this difference ranges from
$0.1-4.98$ per cent,
whereas for a frequency of $3Hz$ it varies in the interval
$0.1-22.78$ per cent. This means that stability is larger in about
23   per cent
for the latter case, a change that is significant in bubble reactors
\cite{joshi}.

The quantities $\beta \equiv \left\{ U_{G_{0}},c_{0},c^{\pm },\tau
_{\varepsilon }\right\} $ and $\beta ^{p}\equiv \left\{ \lambda
^{-1}U_{G_{0}},\lambda ^{-1}c_{0},\lambda ^{-1}c^{\pm }\right\} $ are
plotted as functions of $\varepsilon $ in Fig. 2. The curve for $c_{0}$ is
always between that for $c^{+p}$ and $c^{-p}$. According to the Whitham
stability criterion \cite{with}, when $c_{0}<$ $c^{-}$ the uniform flow is
unstable. This occurs for both distributions, however, for the monodispersed
case it occurs for $\varepsilon >0.353$, whereas for the polydispersed case
the system is stable up to a larger value of the void fraction, e.g. $%
\varepsilon >0.36$. 
\begin{figure}
\includegraphics{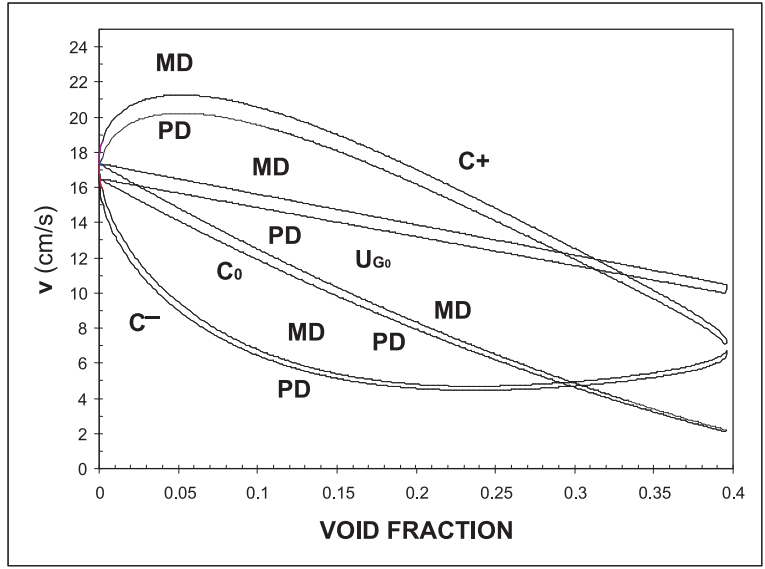}
\caption{$\beta $ and $\beta ^{p}$ as functions of the void fraction 
$\varepsilon $ for the same parameter values as in Fig. 1.}
\end{figure}

\section{DISCUSSION}

Summarizing, in this work we have analyzed the effects of size
polydispersity in several features of the void fraction waves and their
stability properties. We found that the presence of a size distribution
reinforces the stability of the waves, as shown in Figs. 1 and 2.
Furthermore, the per cent difference may be quantified by estimating $\Gamma
\equiv $ $\left\vert \beta -\beta ^{p}\right\vert /\beta $, $\Gamma $ amounts
to a maximum percentual difference of $4.9\%$. 

It is convenient to emphasize once again, that the hydrodynamic
model used in this work \cite{byg} is idealized in many aspects. For
instance, compressibility and hydrodynamic interactions between bubbles and
with the boundaries, have not been taken into account. However, given the
complexity of these effects and of the system itself, the simple dimensional
model proposed by Biesheuvel and Gorissen seems to be a good first step in
modeling the complex behavior of a bubble column. It also ilustrates how
some of the methodology and concepts of kinetic theory and statistical
mechanics may be used to deal with complex phenomena in engineering systems.

We should also mention that in this work we have assumed an initial
polydisperse size distribution and the coalescence of bubbles has not been
considered \cite{salinas}. however, this remains to be assessed. Some other
important effects remain to be considered as well, like the bubble-bubble
interaction mechanisms.Nevertheless the approach followed here by including
the influence of the distribution through the drag effects, considering a
mean field approach,is an attempt to set a first framework for the bubble
size distribution incorporation to further studies. Our results
reinforce this point of view in the sense that a description of a bubble
column based on the concept of randomness of a bubble cloud and average
properties of the fluid motion, may be a useful approach that has not been
exploited in engineering systems.

\begin{acknowledgments}
RFR acknowledges partial financial support from grant DGAPA-UNAM 112503.
\end{acknowledgments}

\end{document}